# Bound state of heavy quarks using a general polynomial potential

## Hesham Mansour* and Ahmed Gamal

Physics Department, faculty of science, Cairo University, Giza, Egypt
Correspondence, Ahmed Gamal, Department of Physics, Cairo University, Giza, Egypt.
E-mail: alnabolci2010@gmail.com
* FInstP

## Abstract
In the present work, the mass spectra of the bound states of heavy quarks $c\bar{c}$, $b\bar{b}$, and $B_c$ meson are studied within the framework of the non-relativistic Schrodinger's equation . First, we solve Schrodinger's equation with a general polynomial potential by Nikiforov Uvarov (NU) method. The energy eigenvalues for any L-value is presented for a special case of the potential. The results obtained are in good agreement with the experimental data and are better than previous theoretical studies.

## 1-Introduction
 In quark and anti-quark system, the quantitively description is given by (quantum chromodynamics (QCD) spectroscopy and the standard model theory) **[1-4]** is important to specify the mechanism of that system and its nature of being bound systems.
The Schrödinger's equation describes quarkonium systems **[5-9]** with a heavy quark and anti-quark interaction (two-body problem). We solve the Schrödinger equation in a spherical-symmetric coordinate and using radial potentials which can be described by the asymptotic limits of QCD which has been qualitatively verified by Lattice QCD calculations **[2-4].**
The main purpose of this paper is that an interaction potential in the quark-antiquark bound system is taken as a general polynomial to get the general eigenvalue and eigenfunction solution then choosing a specific potential according to the description of the physical mechanism of the system**.**

The Nikiforov-Uvarov (NU) method **[10-15]**, gives asymptotic expressions for the eigenfunctions and eigenvalues of the Schrödinger's equation. Hence one can calculate the energy eigenvalues and eigenstates for the spectrum of the quarkonium systems **[17-26]**.

## 2-The Schrödinger equation with polynomial potential

The Schrödinger equation reads

$$\frac{d^2Q}{dr^2} + \left[\frac{2\mu}{\hbar^2}(E - V(r)) - \frac{l(l+1)}{r^2}\right]Q = 0 \tag{1}$$

We use the generalized potential

$$V(r) = \sum_{m=-2}^{m} A_m \, r^m \quad , \quad m = -2, -1, 0, 1, 2, 3, 4, \ldots \tag{2}$$

$$\sum_{m=-2}^{m} A_m \, r^m = A_{-2}\,r^{-2} + A_{-1}\,r^{-1} + A_0 + A_1\,r^1 + A_2\,r^2 + A_3\,r^3 + A_4\,r^4 + \cdots \tag{3}$$

By substituting in Schrödinger equation (1), we get

$$\frac{d^2Q}{dr^2} + \left[\frac{2\mu}{\hbar^2}\left(E - \sum_{m=-2}^{m} A_m\, r^m\right) - \frac{l(l+1)}{r^2}\right]Q = 0 \tag{4}$$

$$\frac{d^2Q}{dr^2} + \left[\sum_{m=-2}^{m} (a,b)_m \, r^m\right]Q = 0 \tag{5}$$

Where $\sum_{m=-2}^{m} (a,b)_m \, r^m = -[b_{-2}r^{-2} + a_{-1}r^{-1} + b_0 + a_1 r^1 + a_2 r^2 + a_3 r^3 + \cdots]$ (6)

and $\quad \frac{2\mu}{\hbar^2} A_m = a_m \ , b_{-2} = l(l+1) + a_{-2} \quad , \quad b_0 = a_0 - \frac{2\mu}{\hbar^2}E = a_0 - \epsilon_0 \tag{7}$

Let $r = \frac{1}{x}$, and substituting in equation(5), we get

$$x^4 \frac{d^2Q}{dx^2} + 2x^3 \frac{dQ}{dx} + \left[-b_{-2}x^2 - a_{-1}x - b_0 - \sum_{m=1}^{m} a_m \left(\frac{1}{x}\right)^m\right]Q = 0 \tag{8}$$

Because of singularity, we expand the polynomial function by Taylor's series around x=0
Let $x = y + \delta$

$$\sum_{m=1}^{m} a_m \left(\frac{1}{x}\right)^m = \sum_{m=1}^{m} \frac{a_m}{(y+\delta)^m} = \sum_{m=1}^{m} \frac{a_m}{\delta^m}\left[1 + \frac{y}{\delta}\right]^{-m} \tag{9}$$

$$\text{let} \quad f(y) = \left[1 + \frac{y}{\delta}\right]^{-m} \tag{10}$$

Neglecting higher order terms, we get

$$f(y) = \left[1 - \frac{m}{\delta}y + \frac{m(m+1)}{2\delta^2}y^2\right] \tag{11}$$

By substituing $x - \delta = y$, we get

$$f(x) = \left[\left[\frac{(m+2)(m+1)}{2}\right] - \left[\frac{m(m+2)}{\delta}\right]x + \frac{m(m+1)}{2\delta^2}x^2\right] \tag{12}$$

By substituting in equation (9), we obtain

$$\sum_{m=1}^{m} a_m \left(\frac{1}{x}\right)^m = \sum_{m=1}^{m} \frac{a_m}{\delta^m}\left[1 + \frac{y}{\delta}\right]^{-m}$$

$$\cong \sum_{m=1}^{m} \frac{a_m}{\delta^m}\left[\left[\frac{(m+2)(m+1)}{2}\right] - \left[\frac{m(m+2)}{\delta}\right]x + \frac{m(m+1)}{2\delta^2}x^2\right] \tag{13}$$

$$\sum_{m=1}^{m} a_m \left(\frac{1}{x}\right)^m \cong \sum_{m=1}^{m} \left[\frac{(m+2)(m+1)a_m}{2\delta^m} - \frac{m(m+2)a_m}{\delta^{m+1}}x + \frac{m(m+1)a_m}{2\delta^{m+2}}x^2\right] \tag{14}$$

By substituting in equation (8), we get

$$x^4 \frac{d^2Q}{dx^2} + 2x^3 \frac{dQ}{dx} + \left[-b_{-2}x^2 - a_{-1}x - b_0 \right.$$
$$\left. - \sum_{m=1}^{m}\left[\frac{(m+2)(m+1)a_m}{2\delta^m} - \frac{m(m+2)a_m}{\delta^{m+1}}x + \frac{m(m+1)a_m}{2\delta^{m+2}}x^2\right]\right]Q = 0 \tag{15}$$

We rearrange equation (15), and divide by $x^4$ where $x \neq 0$ to obtain,

$$\frac{d^2Q}{dx^2} + \frac{2x}{x^2}\frac{dQ}{dx} + \frac{1}{x^4}\left[-\left(b_0 + \sum_{m=1}^{m}\frac{(m+2)(m+1)a_m}{2\delta^m}\right) + \left(\sum_{m=1}^{m}\frac{m(m+2)a_m}{\delta^{m+1}} - a_{-1}\right)x \right.$$
$$\left. - \left(b_{-2} + \sum_{m=1}^{m}\frac{m(m+1)a_m}{2\delta^{m+2}}\right)x^2\right]Q = 0 \tag{16}$$

Let $\left(b_0 + \sum_{m=1}^{m}\frac{(m+2)(m+1)a_m}{2\delta^m}\right) = q \quad , \quad \left(\sum_{m=1}^{m}\frac{m(m+2)a_m}{\delta^{m+1}} - a_{-1}\right)$

$$= w \quad , \quad \left(b_{-2} + \sum_{m=1}^{m}\frac{m(m+1)a_m}{2\delta^{m+2}}\right) = z \tag{17}$$

Equation (16) becomes

$$\frac{d^2Q}{dx^2} + \left(\frac{2x}{x^2}\right)\frac{dQ}{dx} + \frac{1}{x^4}[-q + wx - zx^2]Q = 0 \tag{18}$$

We use the Nikiforov-Uvarov (NU) method **[12, 14]** as mentioned before,

$\tilde{\tau}, \sigma, \tilde{\sigma}, \pi(x), k, \tau(x), \Delta, \lambda, \lambda_n, \rho(x), \Psi(r), \varphi(r), Y(r)$ are symbols used in the Nikiforov-Uvarov (NU) method.

$$\tilde{\tau} = 2x \quad , \quad \sigma = x^2 \quad , \quad \tilde{\sigma} = -q + wx - zx^2 \tag{19}$$

$$\pi(x) = \frac{2x - 2x}{2} \pm \sqrt{\left(\frac{2x - 2x}{2}\right)^2 + q - xw + x^2z + x^2k} \tag{20}$$

$$\pi(x) = \pm\sqrt{q - xw + x^2z + x^2k} \tag{21}$$

We chose the value of $k$ which make the square root in equation (21) as a quadratic term

$$\Delta = b^2 - 4ac = 0 \quad , \quad w^2 - 4(k+z)q = 0 \tag{22}$$

$$(k+z) = \frac{w^2}{4q} \quad \rightarrow \quad k = \frac{w^2}{4q} - z \tag{23}$$

Substituting in equation (21), we get

$$\pi(x) = \pm\sqrt{\left(\frac{w^2}{4q}\right)x^2 - wx + q} \tag{24}$$

$$\pi(x) = \pm\sqrt{\left(\frac{w}{2\sqrt{q}}x - \sqrt{q}\right)^2} \tag{25}$$

We take the negative value

$$\pi(x) = -\left(\frac{w}{2\sqrt{q}}x - \sqrt{q}\right) = -\frac{w}{2\sqrt{q}}x + \sqrt{q} \tag{26}$$

$$\tau(x) = \tilde{\tau}(x) + 2\pi(x) \tag{27}$$

$$\tau(x) = \left[2 - \frac{w}{\sqrt{q}}\right]x + 2\sqrt{q} \quad \text{and} \quad \frac{w}{\sqrt{q}} > 2 \tag{28}$$

$$\lambda = k + \dot{\pi}(x) = \frac{w^2}{4q} - z - \frac{w}{2\sqrt{q}} = \frac{w^2}{4q} - \frac{w}{2\sqrt{q}} - z \tag{29}$$

$$\lambda_n = -n\dot{\tau}(x) - \frac{n(n-1)}{2}\ddot{\sigma}(x) \tag{30}$$

$$\lambda_n = -n\left[2 - \frac{w}{\sqrt{q}}\right] - n(n-1) = -2n + \frac{w}{\sqrt{q}}n - n^2 + n \qquad (31)$$

By equating equations (29) and (31), we get the eigen value equation

$$\lambda_n = \lambda = \frac{w^2}{4q} - \frac{w}{2\sqrt{q}} - z = -n + \frac{w}{\sqrt{q}}n - n^2 \qquad (32)$$

$$\left[\frac{w}{2\sqrt{q}} - \left(n + \frac{1}{2}\right)\right]^2 = z + \frac{1}{4} \quad \rightarrow \quad \frac{w}{2\sqrt{q}} = \sqrt{z + \frac{1}{4}} + \left(n + \frac{1}{2}\right) \qquad (33)$$

$$q = \left[\frac{w}{2n + 1 + 2\sqrt{z + \frac{1}{4}}}\right]^2 \qquad (34)$$

By substituting equation (17) in equation (34), we get

$$b_0 + \sum_{m=1}^{m} \frac{(m+2)(m+1)a_m}{2\delta^m} = \left[\frac{\sum_{m=1}^{m} \frac{m(m+2)a_m}{\delta^{m+1}} - a_{-1}}{2n + 1 + 2\sqrt{b_{-2} + \sum_{m=1}^{m} \frac{m(m+1)a_m}{2\delta^{m+2}} + \frac{1}{4}}}\right]^2 \qquad (35)$$

By substituting equation (7) in equation (35), we get

$$\epsilon_0 = a_0 + \sum_{m=1}^{m} \frac{(m+2)(m+1)a_m}{2\delta^m} - \left[\frac{\sum_{m=1}^{m} \frac{m(m+2)a_m}{\delta^{m+1}} - a_{-1}}{2n + 1 + 2\sqrt{b_{-2} + \sum_{m=1}^{m} \frac{m(m+1)a_m}{2\delta^{m+2}} + \frac{1}{4}}}\right]^2 \qquad (36)$$

Putting $\delta = \frac{1}{r_0}$ and substituting in equation (36), we obtain

$$\epsilon_0 = a_0 + \frac{1}{2}\sum_{m=1}^{m}(m+2)(m+1)\, a_m\, r_0^m$$

$$- \left[\frac{\sum_{m=1}^{m} m(m+2)\, a_m r_0^{m+1} - a_{-1}}{2n + 1 + 2\sqrt{l(l+1) + a_{-2} + \frac{1}{2}\sum_{m=1}^{m} m(m+1)a_m r_0^{m+2} + \frac{1}{4}}}\right]^2 \qquad (37)$$

So, the total energy eigen value is

$$E = A_0 + \frac{1}{2}\sum_{m=1}^{m}(m+2)(m+1)A_m r_0^m$$
$$-\frac{2\mu}{\hbar^2}\left[\frac{\sum_{m=1}^{m}m(m+2)A_m r_0^{m+1} - A_{-1}}{2n+1+2\sqrt{\left(l+\frac{1}{2}\right)^2 + \frac{2\mu}{\hbar^2}A_{-2} + \frac{2\mu}{\hbar^2}\sum_{m=1}^{m}\frac{m(m+1)}{2}A_m r_0^{m+2}}}\right]^2 \quad (38)$$

To find the eigenfunctions for the general potential form

$$Q(x) = \varphi(x)Y(x) \quad (39)$$

First, we calculate $\varphi(x)$

$$\frac{1}{\varphi(x)}\frac{d\varphi(x)}{dx} = \frac{\pi(x)}{\sigma(x)} \quad (40)$$

$$\int \frac{d\varphi(x)}{\varphi(x)} = \int \left[-\frac{\frac{w}{2\sqrt{q}}}{x} + \frac{\sqrt{q}}{x^2}\right]dx \quad (41)$$

$$\varphi(x) = x^{-\frac{w}{2\sqrt{q}}} e^{-\frac{\sqrt{q}}{x}} \quad (42)$$

Second, we calculate $Y(x)$

$$\frac{1}{\rho(x)}\frac{d\rho(x)}{dx} = \frac{\tau(x) - \dot{\sigma}(x)}{\sigma(x)} \quad (43)$$

$$\int \frac{d\rho(x)}{\rho(x)} = \int \left[-\frac{\frac{w}{\sqrt{q}}}{x} + \frac{2\sqrt{q}}{x^2}\right]dx \quad (44)$$

$$\rho(x) = x^{-\frac{w}{\sqrt{q}}} e^{-\frac{2\sqrt{q}}{x}} \quad (45)$$

$$Y(x) = Y_n(x) = \frac{j_{nl}}{\rho(x)}\frac{d^n}{dx^n}[\sigma^n(x)\rho(x)] \quad (46)$$

$$Y(x) = Y_n(x) = j_n x^{\frac{w}{\sqrt{q}}} e^{\frac{2\sqrt{q}}{x}}\frac{d^n}{dx^n}\left[x^{2n-\frac{w}{\sqrt{q}}} e^{-\frac{2\sqrt{q}}{x}}\right] \quad (47)$$

By substituting in equation (39), we obtain

$$Q(x) = \varphi(x)Y(x) = N_n x^{\frac{w}{2\sqrt{q}}} e^{\frac{\sqrt{q}}{x}}\frac{d^n}{dx^n}\left[x^{2n-\frac{w}{\sqrt{q}}} e^{-\frac{2\sqrt{q}}{x}}\right] \quad (48)$$

Where $N_{nl}$ is a normalization constant.

So, the radial wavefunction of Schrödinger equation (1) for a polynomial potential is

$$Q(r) = N_n \, (-1)^n \, n! \; r^{-\frac{1}{2}-\sqrt{z+\frac{1}{4}}} \, e^{\sqrt{q}\,r} \sum_{k=1}^{n} \frac{a_k}{k!} \, r^k \, \frac{d^k}{dr^k}\left[ r^{1+2\sqrt{z+\frac{1}{4}}} \, e^{-2\sqrt{q}\,r} \right] \qquad (49)$$

To find the normalization constant

$$\int_0^\infty |Q(r)|^2 \, r^2 \, dr = 1 \qquad (50)$$

The angular part of the spherical symmetric potential is

$$Y_{lm}(\theta,\varphi) = (-1)^{|m|} \sqrt{\frac{(2l+1)(l-|m|)!}{4\pi\,(l+|m|)!}} \, P_{lm}(\cos\theta) \, e^{im\varphi} \qquad (51)$$

So, the total wavefunction in spherical symmetric potential is

$$\text{Ж}_{nlm}(r,\theta,\varphi) = Q(r)_{nl} Y_{lm}(\theta,\varphi)$$

$$= (-1)^{n+|m|} \, N_n \, n! \sqrt{\frac{(2l+1)(l-|m|)!}{4\pi(l+|m|)!}} \sum_{k=1}^{n} \frac{a_k}{k!} \, r^{\left[k-\frac{3}{2}-\sqrt{z+\frac{1}{4}}\right]} e^{\sqrt{q}\,r} \frac{d^k}{dr^k}\left[ r^{1+2\sqrt{z+\frac{1}{4}}} e^{-2\sqrt{q}\,r} \right] P_{lm}(\cos\theta) \, e^{im\varphi} \quad (52)$$

In our interquark potential, and using the natural units

$$V(r) = \frac{b}{r} + ar + dr^2 + pr^4 \qquad (53)$$

Putting $A_1 = a$, $A_2 = d$,

$\qquad A_4 = P$, $A_{-1} = b$ and the rest of equation (3) equals zero $\qquad (54)$

The energy eigen values equation according to such potentials become

$$E = 3ar_0 + 6dr_0^2 + 15Pr_0^4$$

$$- 2\mu \left[ \frac{3ar_0^2 + 8dr_0^3 + 24Pr_0^5 - b}{2n + 1 + 2\sqrt{\left(l+\frac{1}{2}\right)^2 + 2\mu * [ar_0^3 + 3dr_0^4 + 10Pr_0^6]}} \right]^2 \qquad (55)$$

## 3-Results and Discussion

In this section, we will calculate the spectra for the bound states of heavy quarks such as charmonium, bottomonium and $B_C$ meson. The mass spectra equation is

$$M = m_q + m_{\bar{q}} + E \tag{56}$$

By substituting equation (55) in equation (56), we get

$$M = m_q + m_{\bar{q}} + 3ar_0 + 6dr_0^2 + 15Pr_0^4$$

$$- 2\mu \left[ \frac{3ar_0^2 + 8dr_0^3 + 24Pr_0^5 - b}{2n + 1 + 2\sqrt{\left(l + \frac{1}{2}\right)^2 + 2\mu * [ar_0^3 + 3dr_0^4 + 10Pr_0^6]}} \right]^2 \tag{57}$$

Equation (57) depends on the potential parameters $(a, b, d, p \text{ and } r_0)$ which will be obtained from the experimental data.

In the case of charmonium $[\Psi = c\bar{c}]$, the rest mass equation is

$$M = 2.55 + 3ar_0 + 6dr_0^2 + 15Pr_0^4$$

$$- 2\mu \left[ \frac{3ar_0^2 + 8dr_0^3 + 24Pr_0^5 - b}{2n + 1 + 2\sqrt{\left(l + \frac{1}{2}\right)^2 + 2\mu * [ar_0^3 + 3dr_0^4 + 10Pr_0^6]}} \right]^2 \tag{58}$$

And we get the following masses in GeV

The charmonium system $r_0 = 0.9302$, $a = 25.6245$, $b = -7.11$, $d = -28.7354$, $p = 7.1026$

| n | L | type | Present | Ref (27) | Ref (28) | Ref (29) | Ref (30) | Ref (31) | PDG (25) |
|---|---|------|---------|----------|----------|----------|----------|----------|----------|
| 1 | 0 | 1S | 2.98301 | 2.989 | 2.981 | 2.984 | 2.925 | 2.979 | 2.984 |
| 2 | 0 | 2S | 3.67657 | 3.681 | 3.635 | 3.679 | 3.676 | 3.673 | 3.686 |
| 3 | 0 | 3S | 4.01076 | 4.129 | 4.039 | 4.030 | 3.803 | 4.022 | 4.039 |
| 4 | 0 | 4S | 4.19694 | 4.514 | 4.427 | 4.281 | - | 4.273 | 4.421 |
| 5 | 0 | 5S | 4.31118 | 4.799 | 4.811 | 4.459 | - | 4.446 | -- |
| 6 | 0 | 6S | 4.38627 | 5.124 | 5.155 | - | - | 4.595 | -- |
| 1 | 1 | 1P | 3.40565 | 3.428 | 3.413 | 3.415 | 3.323 | 3.433 | 3.415 |
| 2 | 1 | 2P | 3.87319 | 3.955 | 3.870 | 3.848 | 3.833 | 3.937 | 3.927 |
| 3 | 1 | 3P | 4.11774 | 4.296 | 4.301 | 4.146 | - | 4.131 | -- |
| 4 | 1 | 4P | 4.26148 | 4.653 | 4.698 | - | - | - | -- |
| 5 | 1 | 5P | 4.35306 | 4.983 | - | - | - | - | -- |
| 1 | 2 | 1D | 3.79208 | 3.755 | 3.813 | 3.808 | 3.869 | 3.799 | - |
| 2 | 2 | 2D | 4.07279 | 4.176 | 4.220 | 4.112 | 3.806 | 4.103 | -- |
| 3 | 2 | 3D | 4.23402 | 4.549 | 4.574 | 4.340 | - | 4.331 | -- |
| 4 | 2 | 4D | 4.33508 | 4.89 | 4.920 | - | - | - | -- |
| 1 | 3 | 1F | 4.04628 | 3.99 | 4.041 | - | - | - | - |
| 2 | 3 | 2F | 4.21806 | 4.378 | 4.361 | - | - | - | -- |
| 3 | 3 | 3F | 4.32474 | 4.73 | - | - | - | - | -- |

In the following we draw the radial wave functions as a function of the radius in figure (1),

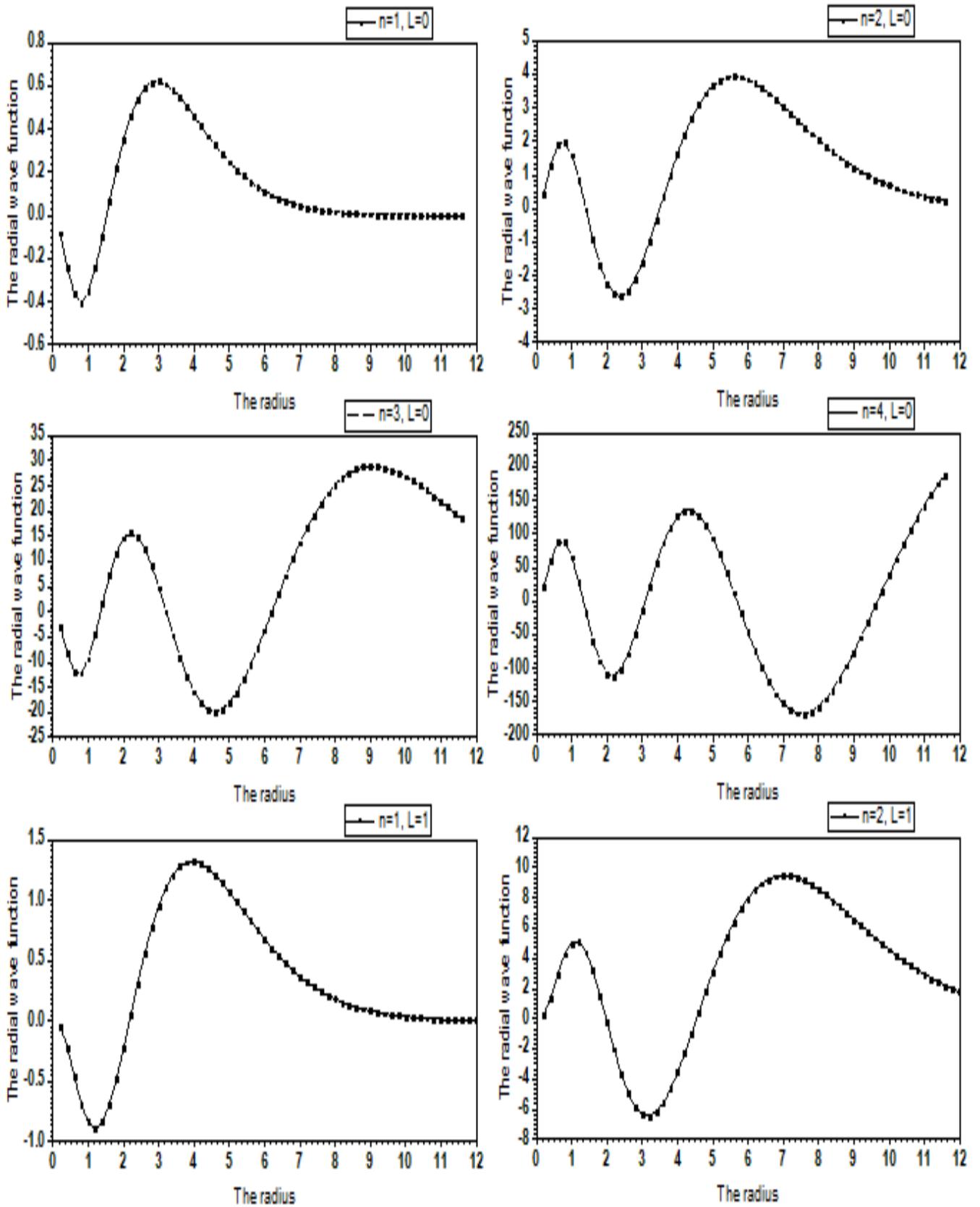

Figure 1, graphs represent the relation between the radial wave function and the radius in different n, L states according to experimental energy states in *the charmonium system*

In the case of bottomonium $[\Upsilon = b\bar{b}]$, the rest mass equation is

$$M = 8.842 + 3ar_0 + 6dr_0^2 + 15Pr_0^4 - 2\mu \left[ \frac{3ar_0^2 + 8dr_0^3 + 24Pr_0^5 - b}{2n + 1 + 2\sqrt{\left(l + \frac{1}{2}\right)^2 + 2\mu * [ar_0^3 + 3dr_0^4 + 10Pr_0^6]}} \right]^2 \quad (59)$$

And we get the following masses in GeV

The bottomonium system $r_0 = 0.8313, a = 8.2234, b = -2.843, d = -8.8278, p = 2.532$

| n | L | type | Present | Ref (27) | Ref (28) | Ref (30) | Ref (29) | Ref (32) | PDG (25) |
|---|---|------|---------|----------|----------|----------|----------|----------|----------|
| 1 | 0 | 1S | 9.39858 | 9.428 | 9.398 | 9.390 | 9.414 | 9.389 | 9.398 |
| 2 | 0 | 2S | 10.05295 | 9.979 | 10.023 | 10.015 | 10.089 | 10.016 | 10.023 |
| 3 | 0 | 3S | 10.35403 | 10.359 | 10.355 | 10.343 | 10.327 | 10.351 | 10.355 |
| 4 | 0 | 4S | 10.51698 | 10.683 | 10.586 | 10.597 | - | 10.611 | 10.579 |
| 5 | 0 | 5S | 10.61499 | 10.975 | 10.869 | 10.811 | - | 10.831 | 10.876 |
| 6 | 0 | 6S | 10.67848 | 11.243 | 11.088 | 10.988 | - | 11.023 | 11.019 |
| 1 | 1 | 1P | 9.83834 | 9.806 | 9.859 | 9.864 | 9.815 | 9.865 | 9.859 |
| 2 | 1 | 2P | 10.24859 | 10.205 | 10.233 | 10.220 | 10.254 | 10.226 | 10.232 |
| 3 | 1 | 3P | 10.45758 | 10.54 | 10.521 | 10.490 | - | 10.502 | -- |
| 4 | 1 | 4P | 10.57828 | 10.84 | 10.781 | - | - | 10.732 | -- |
| 5 | 1 | 5P | 10.65423 | 11.115 | - | - | - | 10.933 | -- |
| 1 | 2 | 1D | 10.19392 | 10.075 | 10.161 | 10.153 | 10.145 | 10.151 | 10.163 |
| 2 | 2 | 2D | 10.42783 | 10.423 | 10.449 | 10.436 | - | 10.442 | -- |
| 3 | 2 | 3D | 10.56034 | 10.733 | - | - | - | 10.680 | -- |
| 4 | 2 | 4D | 10.64259 | 11.015 | - | - | - | 10.886 | -- |
| 1 | 3 | 1F | 10.4114 | 10.283 | 10.343 | 10.338 | - | - | - |
| 2 | 3 | 2F | 10.55055 | 10.604 | 10.610 | - | - | - | -- |
| 3 | 3 | 3F | 10.6363 | 10.894 | - | - | - | - | -- |

In the following we draw the radial wave functions as a function of the radius in figure (2) and figure (3)

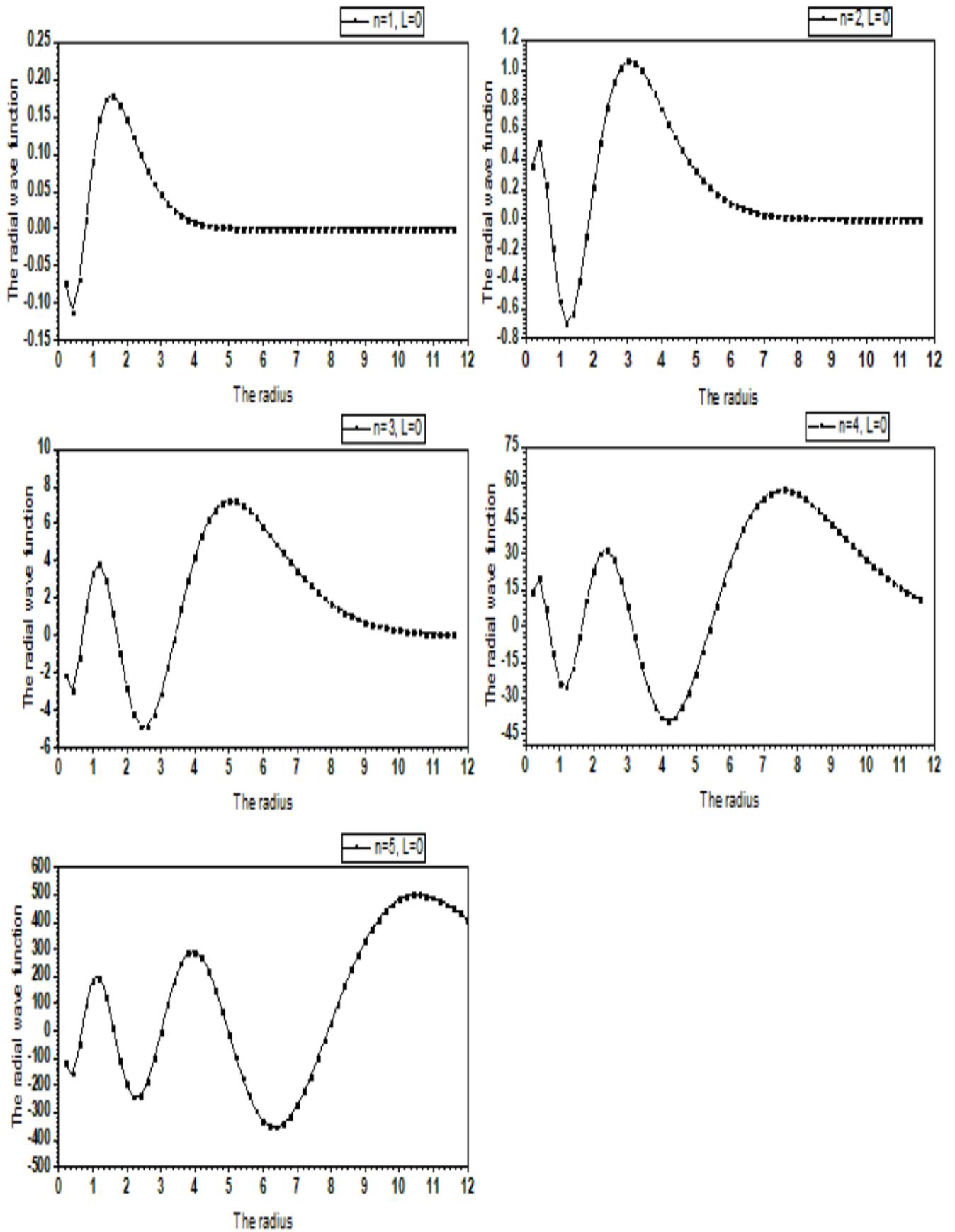

Figure 2, graphs represent the relation between the radial wave function and the radius in different n, L according to experimental energy states in *the bottomonium system*

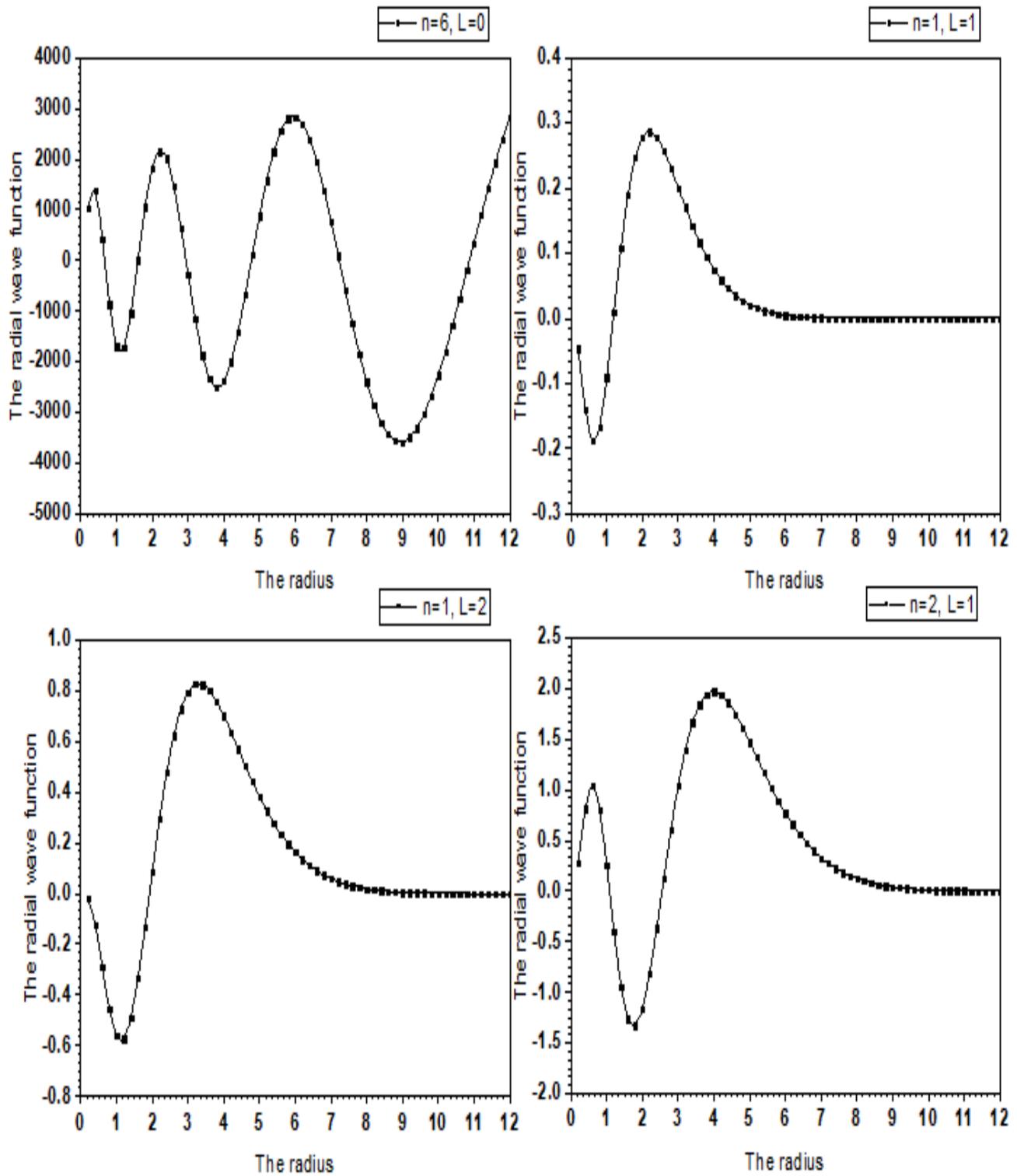

Figure 3, graphs represent the relation between the radial wave function and the radius in different n, L according to experimental energy states in the bottomonium system

In the case of the meson $[B_c = b\bar{c}]$, the rest mass equation is

$$M = 5.696 + 3ar_0 + 6dr_0^2 + 15Pr_0^4$$

$$- 2\mu \left[ \frac{3ar_0^2 + 8dr_0^3 + 24Pr_0^5 - b}{2n + 1 + 2\sqrt{\left(l + \frac{1}{2}\right)^2 + 2\mu * [ar_0^3 + 3dr_0^4 + 10Pr_0^6]}} \right]^2 \quad (60)$$

And we get the following masses in GeV

The meson $B_c$ system $r_0 = 0.2235$, $a = 2.2486$, $b = -2.5912$, $d = -0.17658$, $p = 4.2183$

| n | L | Type | Present | Ref (27) | Ref (28) | Ref (33) | Ref (34) | Ref (35) | PDG (25) |
|---|---|------|---------|----------|----------|----------|----------|----------|----------|
| 1 | 0 | 1S | 6.27543 | 6.272 | 6.272 | 6.278 | 6.271 | 6.275 | 6.275 |
| 2 | 0 | 2S | 6.84121 | 6.864 | 6.842 | 6.863 | 6.855 | 6.838 | 6.842 |
| 3 | 0 | 3S | 7.04335 | 7.306 | 7.226 | 7.244 | 7.250 | - | - |
| 4 | 0 | 4S | 7.13794 | 7.684 | 7.585 | 7.564 | - | - | - |
| 5 | 0 | 5S | 7.18967 | 8.025 | 7.928 | 7.852 | - | - | -- |
| 6 | 0 | 6S | 7.22101 | 8.340 | - | 8.120 | - | - | -- |
| 1 | 1 | 1P | 6.83016 | 6.686 | 6.699 | 6.748 | 6.706 | 6.672 | - |
| 2 | 1 | 2P | 7.03865 | 7.146 | 7.094 | 7.139 | 7.122 | 6.914 | - |
| 3 | 1 | 3P | 7.13552 | 7.536 | 7.474 | 7.463 | - | - | -- |
| 4 | 1 | 4P | 7.18827 | 7.885 | 7.817 | - | - | - | -- |
| 5 | 1 | 5P | 7.22012 | 8.207 | - | - | - | - | -- |
| 1 | 2 | 1D | 7.03763 | 6.990 | 7.029 | 7.026 | 7.045 | 6.980 | - |
| 2 | 2 | 2D | 7.135 | 7.399 | 7.405 | 7.363 | - | - | -- |
| 3 | 2 | 3D | 7.18796 | 7.761 | 7.750 | - | - | - | -- |
| 4 | 2 | 4D | 7.21993 | 8.092 | - | - | - | - | -- |
| 1 | 3 | 1F | 7.13477 | 7.234 | 7.273 | - | 7.269 | - | - |
| 2 | 3 | 2F | 7.18783 | 7.607 | 7.618 | - | - | - | -- |
| 3 | 3 | 3F | 7.21985 | 7.946 | - | - | - | - | -- |

In the following we draw the radial wave functions as a function of the radius in figure (4)

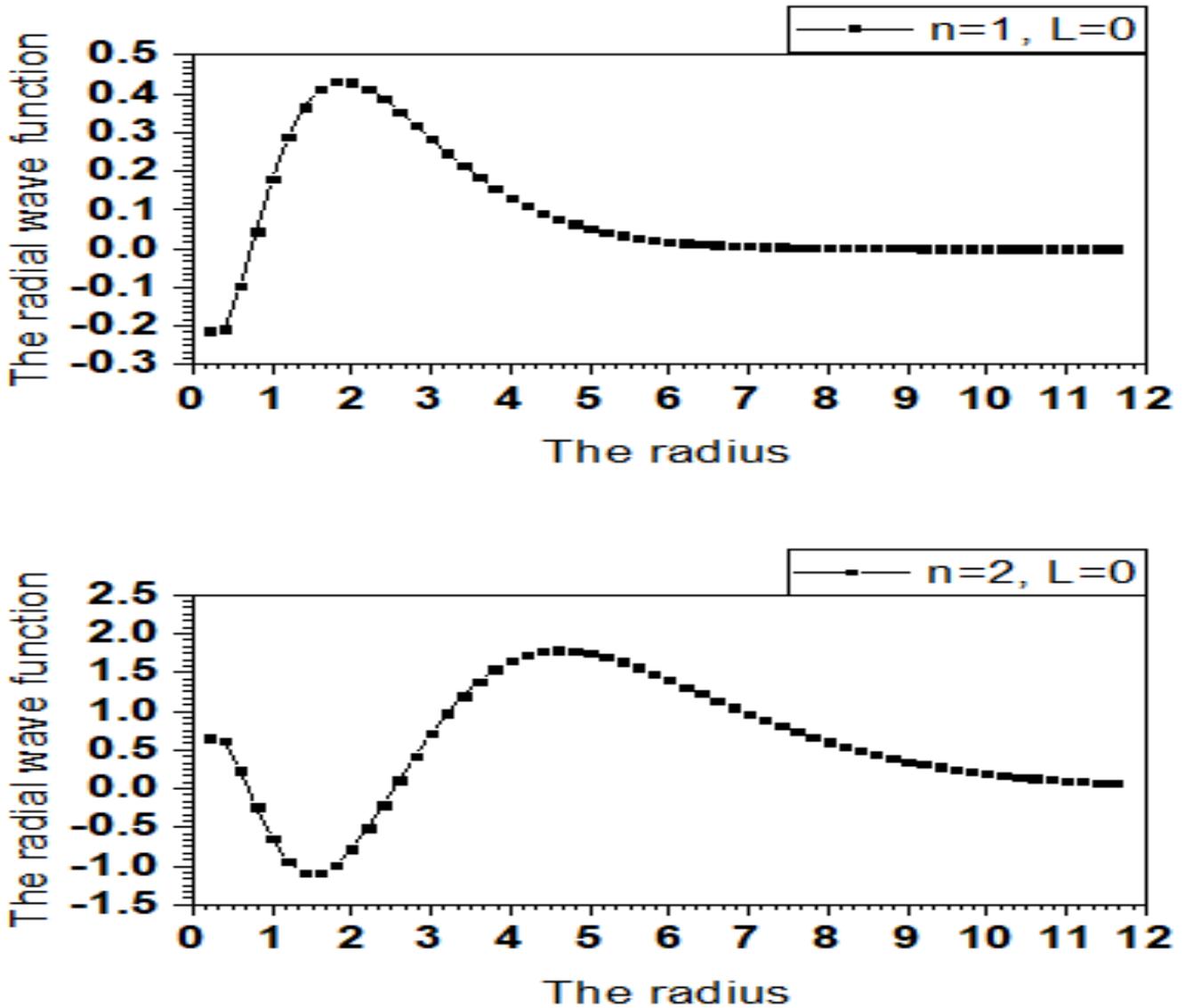

Figure 4, graphs represent the relation between the radial wave function and the radius in different n, L according to experimental energy states in the meson $[B_c = b\bar{c}]$ system

In conclusion, form the tables, we found that our theoretical work is comparable with the experimental data and explains the behavior of the quarkonium systems. The difference between the experimental data and theoretical work may be because we neglect the spin terms, so, the spin can also be considered if one uses relativistic corrections and the appropriate relativistic Schrödinger's equation. From the figures which represent the quarkonium radial state wave functions, one can calculate physical parameters like the decay parameter.